# Ensuring patients' privacy in a cryptographic-based-electronic health records using bio-cryptography

## Adebayo Omotosho*

Department of Computer Science and Information Technology,
Bells University of Technology,
P.M.B 1015, Ota, Ogun State, Nigeria
Email: bayotosho@gmail.com
*Corresponding author

## Justice Emuoyibofarhe

Department of Computer Science and Engineering,
Ladoke Akintola University of Technology,
P.M.B 4000, Ogbomoso, Oyo State, Nigeria
Email: eojustice@gmail.com

## Christoph Meinel

Hasso Plattner Institute (HPI) for IT Systems Engineering,
University of Potsdam,
Potsdam, 14482, Germany
Email: meinel@hpi.de

**Abstract:** Several recent works have proposed and implemented cryptography as a means to preserve privacy and security of patient's health data. Nevertheless, the weakest point of electronic health record (EHR) systems that relied on these cryptographic schemes is key management. Thus, this paper presents the development of privacy and security system for cryptography-based-EHR by taking advantage of the uniqueness of fingerprint and iris characteristic features to secure cryptographic keys in a bio-cryptography framework. The results of the system evaluation showed significant improvements in terms of time efficiency of this approach to cryptographic-based-EHR. Both the fuzzy vault and fuzzy commitment demonstrated false acceptance rate (FAR) of 0%, which reduces the likelihood of imposters gaining successful access to the keys protecting patients' protected health information. This result also justifies the feasibility of implementing fuzzy key binding scheme in real applications, especially fuzzy vault which demonstrated a better performance during key reconstruction.

**Biographical notes:** Adebayo Omotosho received his PhD in Computer Science at Ladoke Akintola University of Technology in 2016. He is a Seasoned Computer Programmer and has taken part in a number of programming competitions in C/C++/C#. His current research interests are health informatics, computer security, big data analytics and biometrics.

Justice Emuoyibofarhe is a Professor of Computing at Ladoke Akintola University of Technology. He received his PhD in 2004. He specialises in neuro-fuzzy computing computational optimisation. He had post-doctoral fellowship at the Centre of Excellence for Mobile e-service, University of Zululand, South Africa in 2006. He is a Member of the IEEE Computational Intelligence Society. He is also a Visiting Researcher at the Hasso Plattner Institute, University of Potsdam, Germany. His present research area is in the application of mobile computing and wireless communication to e-health and telemedicine.

Christoph Meinel is a German Scientist and a University Professor of Computer Sciences. He is President and CEO of the Hasso Plattner Institute (HPI) for IT Systems Engineering at the University of Potsdam (Germany), and a Professor for Internet Technologies and Systems. Besides his teaching activities in Potsdam, he is an Honorary Professor at the Technical University of Beijing (China), a Visiting Professor at the Shanghai University (China), and a Senior Research Fellow of SnT at the University of Luxembourg. He is a Chairman or a member of various international scientific boards and program committees, and has organised several internal symposia and conferences.

## 1  Introduction

Healthcare Information and Management Systems Society defines electronic health record (EHR) as "a secure, real-time, point-of-care, patient centric information resource for clinicians" (HIMSS, 2003). EHR is on the verge of receiving widespread adoption as an instrument for improving the understanding of the state of health of individuals as it contains useful, legal and computerised historical health data from a variety of sources. Over time, a patient's EHR accumulates significant information, such as identifying information, hospital visitations, laboratory data, surgery, radiology reports, allergies, vital signs, immunisations, prescriptions, sexual preference, psychological profiles, physician progress notes and among other relevant data that defines a medical record (Mercuri, 2004; Tiwari and Kumar, 2015). The importance of these data to healthcare providers, patients and cybercriminals made it necessary that the EHR designs be responsible for securing patients' personal data, health records and managing the access rights to them (IMIA, 2000; Brumen et al., 2013; CTI, 2016).

In several studies, patients' participation has been identified as crucial to manage records, because patients' growing concern over the privacy and security of their personal and sensitive data stored in EHR has been a contributing factor to the slow adoption of this technology (Escarfullet et al., 2012; Dinev et al., 2016). Based on a survey report, 73% of information privacy and security issues reported by health organisation over the years involved electronic data (Hewapathiranae, 2011). Studies have shown that millions of patient records have been breached and are at the risk in the hands of malicious users (Robertson, 2014; Harris, 2016). In the USA alone, since 2005, more than 300 data breaches in which 100,000 or more records were compromised have been publicly disclosed in the health industry (Collins and Robertson, 2014). Also, since 2009, the annual number of cyber-attacks against the healthcare sector has drastically increased; often the number of attacks exceeds the previous year's count by at least 40% (ICIT, 2016). Similarly, more often than expected, the majority of threats that healthcare organisations face are internal among other sources of threats (Gordon et al., 2006; Computer Economics, 2007; Srinivasan, 2016; Sanzgiri and Dasgupta, 2016). With inside attacks either continuously outnumbering or being the leading cause of external threats, most research still paid more attention to outsiders. The cost of privacy and confidentiality breaches is very difficult to recover for both the healthcare provider and patient (Appari and Johnson, 2010).

For privacy reasons, patients at some point may be unwilling to disclose important information about some health problems such as psychiatric behaviour and HIV as their violation or disclosure may lead to social stigma, unfair treatment by employers and possibly irreversible damage to their professional reputation (Omotosho and Emuoyibofarhe, 2014; Carroll, 2016). Privacy and security are crucial factors in electronic healthcare; cryptography is the science of information scrambling and it has been recommended to be used everywhere for ensuring privacy and security (Demiris, 2004; Escarfullet et al., 2012; Hewitt, 2013; Bhartiya and Mehrotra, 2015; Grunwel and Sahama, 2016). Through cryptography, a number of autonomous patients controlled health record systems have been proposed and were discussed in Omotosho and Emuoyibofarhe (2014). Cryptographic-based-EHR has become very popular but patients' privacy could still be violated, if in a patient controlled system, encryption keys are not efficiently managed. Likewise, clinician work flow would be hampered, if in the course of ensuring privacy, health data are not available in decrypted form for authorised access.

EHR promises monolithic benefits if the mechanisms tailored towards achieving privacy and security are not too cumbersome and can easily be managed by both the patient and physician without affecting the timeliness, clinician workflow and quality of healthcare service delivery (Hillestad et al., 2005; Omotosho and Emuoyibofarhe, 2014). The focus of this study is to address privacy and security issues by taking advantage of the fuzziness of biometrics data to protect cryptographic keys in the developed EHR application using the concept of bio-cryptography. Bio-cryptography blends biometrics with cryptography by combining the benefits of both technologies to provide a stronger means to protect against system attacks. These techniques protect a secret key using biometric feature or generate a key from biometric features. There are two popular approaches to implement bio-cryptography: the key binding approach – which is the binding of cryptographic keys along with biometric template and the key generating approach – which involves generating cryptographic keys from biometric templates (Stavroulakis and Stamp, 2010; Scheirer et al., 2010; Das, 2012). Bio-cryptography key binding method will be used to control access to patient's key by physicians.

Popular bio-cryptography key binding methods are the fuzzy vault and fuzzy commitment. The archetype for fuzzy vault construct was proposed by Juels and Sudan (2002). Fuzzy vault is an error tolerant encryption that is suited for applications combining both biometrics and cryptography. Fuzzy commitment technique was proposed by Juels and Wattenberg (1999). They combined the error-correcting codes and cryptography to fuzzy commitment scheme (Rathgeb and Uhl, 2011). The purpose of fuzzy commitment scheme is to bind biometric features of a user with a key prepared with an error correction code to overcome the fuzziness of biometric measurements. It is very similar to fuzzy vault except that it is not an error tolerant encryption operation because it does not support order invariance. Several variants of fingerprint fuzzy vault and iris fuzzy commitment have been widely modified and experimented with as presented in our previous work in Omotosho and Emuoyibofarhe (2014). Even though, this work centres on fuzzy vault because it is more suitable for applications using biometrics and cryptography, both techniques were used in this study since neither of the bio-cryptography approach have been applied in real applications like the EHR where timeliness is of high value. This strategy together with the designed joint EHR management architecture has the capability of improving patients' privacy and security in a cryptographic-based-EHR.

## 2  Methodology and system description

The development of the privacy enhanced bio-cryptography key management system involves the steps highlighted in Table 1.

**Table 1**  Summary of research activities

| Steps | Research activities |
|---|---|
| 1 | The formulation of a cryptographic key generation scheme where unique and reproducible keys were derived from image statistics features of grey level cooccurrence matrix |
| 2 | The modification of a fuzzy vault bio-cryptographic key protection technique for protecting the keys produced |
| 3 | The derivation of a shared EHR privacy rule scheme through EHR questionnaires. This ensured that a successful access does not imply privilege to a complete EHR of patient. A requester of a patient EHR cannot view the EHR parts with patients' privacy set |
| 4 | The implementation of an EHR accountability scheme that ensured that insiders' infringements on EHR are traceable. The implementation and evaluation of the EHR system were carried out using tools, such as Microsoft Office Visio, Microsoft Visual C#, Java and Matlab |

### 2.1  Cryptographic key generation method using grey level cooccurrence matrix (GLCM) technique

Unique encryption keys were generated from second order statistical features of image using GLCM and this is similar to the cryptographic key generation method proposed in Omotosho and Emuoyibofarhe (2015). This approach makes it possible to reproduce the same key repeatedly from the same image and patients do not have to store or memorise keys. Unlike image steganography where data is hidden in an image, keys were generated and not hidden in the image and this makes it difficult to steal patient's key. The algorithm was modified to take time attributes which defines the usage limit period of each key generated as shown in Section 2.1.1. The time attributes were presented in the form of days, hours, minutes and seconds. This is to ensure that the definition of privacy is fulfilled at key generation level. That is, the right of a patient to determine when and how their records are accessed. This way, a patient reduces the risk of having encrypted part of EHR available to a physician once consultation is completed.

*2.1.1 Algorithm for key generation (adapted from Omotosho and Emuoyibofarhe (2015))*

1   Start

2   Reading the image into array

3   Convert the image into a matrix [20, *n*]. where *n* is number of column in the image matrix

4   Specify the distance between neighbouring pixels, *r*

5   Specify the angle, *D* of the grey level cooccurrence matrix

    i    If *D* = 90° then set offset = *r*\*[–1 0]

    ii    If *D* = 45° then set offset = *r*\*[–1 1]

    iii    If *D* = 135° then set offset = *r*\*[–1 –1]

    iv    If *D* = 0° then set offset = *r*\*[0 1]

6   Find the trace of the matrix (sum of the diagonal of the matrix)

7   Compute other image descriptor values (Contrast, Correlation, Energy, Entropy, Homogeneity)

8   Set *k* = Trace \* Contrast \* Correlation \* Energy \* Entropy \* Homogeneity

9   Set count = location of decimal point and set GLCMKey = 1.0

10   GLCMKey\* = Math.Pow (10, count)

11   Show time option, Hour, Minutes and Seconds

12   Set active time and expiry time

13   Stop

## 2.2 Bio-cryptography key binding approach

The keys generated were stored in user's biometric templates using two different biometrics. The fingerprint minutiae and iris codes extracted from captured users templates were used in the fuzzy vault and fuzzy commitment key constructs, respectively. The rationales for the biometrics choice are described as follows.

### 2.2.1 Fingerprint

The fingerprint minutiae served as the biometric templates, which were stored in the database in the form of fuzzy vault helper data. The fundamental reasons for the choice of this biometrics in this research are:

i   enrolment takes little time

ii   large number of people are conversant with identification based on the use of fingerprint iii   works well with fuzzy vault.

## 2.2.2 Iris

Iris codes are the biometric templates which are stored in the database. This biometrics is selected for the following reasons:

i       iris is well protected against wear and damage because it is an internal organ

ii      iris texture is stable over a long time; it is one of the most accurate and consistent biometric

iii     it has one of the lowest FRR

iv      works well with fuzzy commitment.

## 2.2.3 Fuzzy vault algorithm

Fuzzy vault relies on polynomial construction and the stronger the polynomial, the more security of the system. The complexity in decoding the vault polynomial construction could result in trading strong security for higher running time and high FRR. The modified algorithm generated a polynomial of degree $n$ and not $n − 1$ for a key of length. This was done to make the polynomial stronger and also rather than projecting the keys as coefficients of the polynomial, new sets of coefficients were generated.

### 2.2.3.1 Encryption algorithm

i       A cryptographic key generated from GLCM is represented as *ABCD* where *A*, *B*, *C* and *D* are integer numbers.

ii      Using *A*, *B*, *C* and *D* as the polynomial root and introducing variable *x* as the unknown, a polynomial can be derived thus:

$$(x - A)(x - B)(x - C)(x - D)$$

$$x^4 - (A + B + C + D)x^3 + (AB + AC + AD + BC + BD + CD)x^3 - (ABC + ABD + ACD + BCD)x + ABCD.$$

iii     Extracting the coefficient of the resulting polynomial

$$1, -(A + B + C + D), (AB + AC + AD + BC + BD + CD),$$
$$-(ABC + ABD + ACD + BCD), \text{ and } ABCD.$$

iv      Converting coefficients ($c_i$) to base64

$$1 = c1; -(A + B + C + D)_{64} = c2; (AB + AC + AD + BC + BD + CD)_{64} = c3;$$
$$-(ABC + ABD + ACD + BCD)_{64} = c4$$

v       Given a biometric data $B = \{b_0 b_1 b_2 b_3 b_4\ b_5 b_6 b_7 b_8 b_9 \ldots, b_n\}$ extracted from a template.

vi      Convert *B* into the same base as the coefficients $c1$, $c2$, $c3$, $c4$, and $c5$.

Given that $0 \leq z \leq n$ where $z$ in a position within length $B(n)$

Fuzzy vault data = $\{b_0 b_1 b_2 b_3 \ldots b_{n-z} c1 b_{n-z} c2 \ldots b_{n-z} c3 \ldots b_{n-z} c4 \ldots b_n\}$.

### 2.2.3.2 Decryption algorithm

i   If biometrics verification was successful, then {c1, c2, c3, c4, and c5} will be extracted.

ii  Convert {*c*1, *c*2, *c*3, *c*4, and *c*5} to decimal {*d*1, *d*2, *d*3, *d*4, and *d*5}.

iii Reconstruct polynomial with coefficients {*d*1, *d*2, *d*3, *d*4, and *d*5} and symbolic variable *x*:

$$d_1x^4 + d_2x^3 + d_3x^2 + d_4x^1 + d_5x^0.$$

iv  Solving the polynomial $d_1x^4 + d_2x^3 + d_3x^2 + d_4x + d_5 = 0$

$$(x-A)(x-B)(x-C)(x-D).$$

### 2.2.4 Fuzzy commitment algorithm

Similar to Hao et al. (2006), Hadamard codes were used with the iris code to correct bit errors resulting from natural biometric variance while burst errors arising from distortions were corrected using Reed-Solomon codes. The algorithm for fuzzy commitment encryption and decryption are presented in Sections 2.2.4.1 and 2.2.4.2.

### 2.2.4.1 Encoding algorithm

i   Given a biometric data $B1 = \{b_0b_1b_2b_3b_4b_5b_6b_7b_8b_9 \ldots b_n\}$ where b are binary representation of iris code.

ii  Apply Reed Solomon and Hadamard ECC on the original biometrics *B*1 to generate a codeword.

iii Let commitment $B2 = \{b_0b_1b_2b_3b_4b_5b_6b_7b_8b_9 \ldots b_n\}$ + key + codeword.

iv  Set hamming distance threshold = 0.30. This is to balance the system's false acceptance rate (FAR) and FRR because matching has to be done using error correction schemes and this prevents the use of complex biometric template matchers developed uniquely for matching the original template.

v   Discard *B*1 and store commitment *B*2 in the database as the fuzzy commitment.

### 2.2.4.2 Decoding algorithm

i   Given a query template $A = \{a_0a_1a_2a_3a_4a_5a_6a_7a_8a_9 \ldots a_n\}$

ii  At authentication **A** is compared *B*2.

iii Key is unlocked if the difference of the hamming distance is less than threshold, which implies that *A* is sufficiently close to the original template.

### 2.3 EHR privacy scheme

A total of 250 questionnaires were administered in 17 different hospitals in Lagos State and Ogun State to identify some attributes of patients EHR which were used to formulate the record partitions. Random sampling technique was used with the questionnaires to generate patient's parameters. The questionnaire was standardised using five level Linkert scale and in order for data to be representable in the EHR, the results were further trimmed down to two scales. The EHR questionnaire acquired data for basic, confidential and emergency sections of patient's health record, with respect to record sharing. Consequently, in the scope of this work:

i    *basic attributes* implies information about patients which should be available at all times to physicians, hospitals and other entities such as employers, organisations and research institutes without acquiring patient's permission ii   *confidential attributes* are information of patient health record that should not always be available or shared with other healthcare providers, employers, organisations or research institutes without patient's full consent iii  *emergency attributes* are attributes that should be sufficient to take care of patients in pre-hospital care or emergency situations.

The result of the privacy scheme was partitioned EHR based on the categories. The presented privacy architecture disabled the locking of certain fields (basic) of EHR determined by some privacy policies to prevent hampering the clinician workflow and practice. This is because allowing patients to lock all their records could be burdensome, time consuming and impractical. From Figure 1, in order to achieve privacy, the system provides a joint EHR – patient privacy rule. EHR privacy rule is enforced by the provider while a patient as depicted in the architecture can use a symbol such as 'H' to have exclusive right to an attributes (confidential) of his or her record. This feature was implemented in the form of patient key generated from the method presented earlier. The components of this architecture are described in the following sub sections.

**Figure 1**    EHR rule and accountability architecture (see online version for colours)

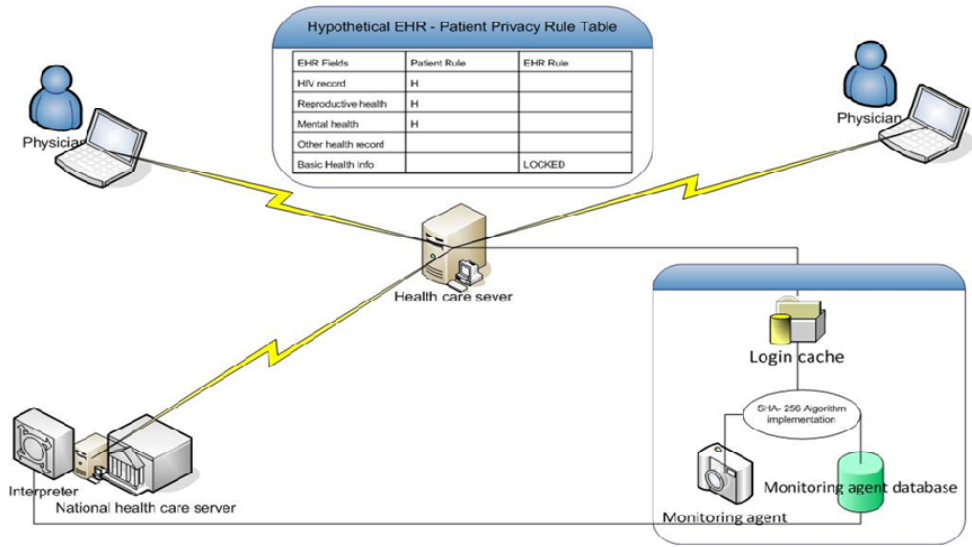

### 2.3.1 Monitoring phase

The monitoring phase involves the process of tracking the access time of records, digital signature of practitioner that access records, operations performed during access and a list of updated changes. After accessing records, SHA-2 hashing algorithm was implemented to create hash of every defined operation. This helps to determine the integrity of the EHR to verify that the original contents have not been changed and if it has then a log is created.

### 2.3.2 Interpreting phase

The interpreting phase is where the hash translator represents every hashed operation performed on the EHR and is generated on the patients' accessed records. It contains all defined operations that can be performed and hash formats. Operations hashed in this study were limited to read and write. This is because they are the most commonly used avenue used by malicious insiders. Other loops such as date of last access of record were also embedded in the system. When records are accessed, the interpreter uses its standard hash format that contains all possible operations that can be performed on records to compare previously hashed values with new values and updates the EHR. All changes are then saved on the national healthcare server.

## 2.4 The overall bio-cryptography key management system

The complete EHR system architecture that combines privacy, security and accountability is shown in Figure 2. The components of the architecture are described as follows:

### 2.4.1 Patient and physician

The focus of the work is mainly to prevent inside attacks, physicians are the primary insiders here, as they are the professionals concerned with the diagnosis and treatment of diseases, injury, physical and mental impairments and maintaining of patient's health. The patients in this system represent any recipient of healthcare services.

### 2.4.2 Images, image analysis, and key

Each unique cryptographic key is generated from any random image supplied by the users. Popular image formats such as the joint photographic experts group (JPEG), bitmap images (BMP) and portable network graphics (PNG) are supported. Images can as well be deleted after use. The image analysis function uses the GLCM to extract statistical texture feature of user specified images. A suitable key for symmetric encryption is generated and the image can then be discarded once used. Users will not have to create lengthy keys from a random number generator or hard type a strong key because the production of keys from images makes it very difficult for inside attackers to know which image was used or how the keys were generated.

**Figure 2**   Developed bio-cryptography key management architecture (see online version for colours)

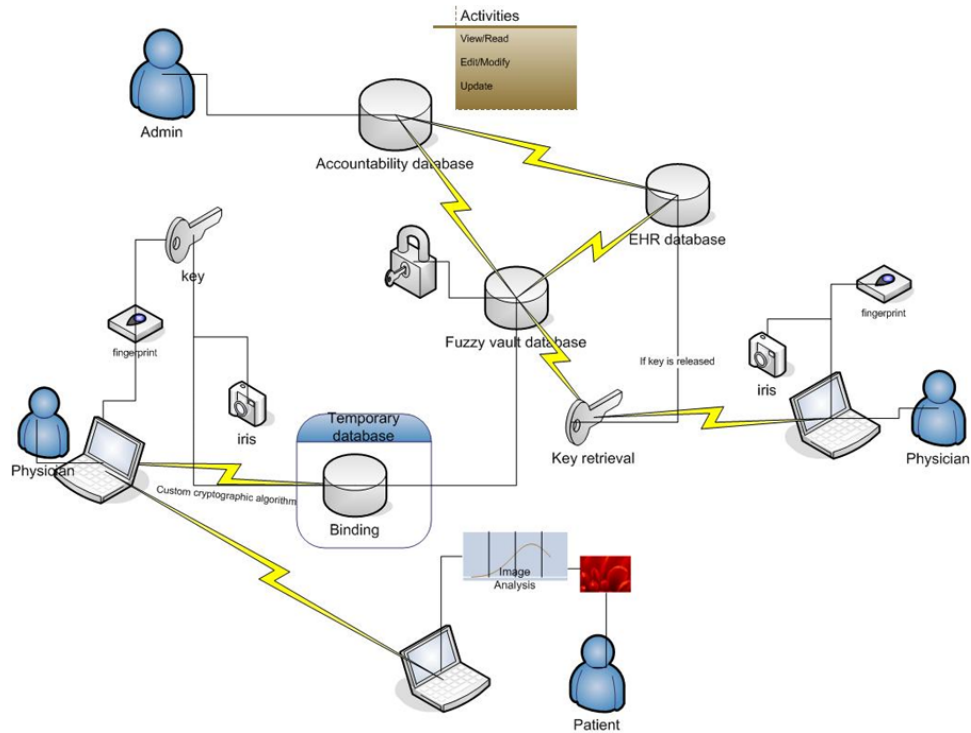

### 2.4.3 Iris scanner and fingerprint scanner

This study makes use of two different biometrics data capturing devices, a dual iris imaging device that captures high quality biometric images of the two irises simultaneously was used. Also, a fingerprint device with high minutiae image quality was used to capture the fingerprint biometrics data. More importantly, these devices have capabilities which are essential in very large scale deployments.

### 2.4.4 Temporary database and fuzzy vault

Once both the keys to be protected and biometrics data are available, they are combined to form the helper data for the bio-cryptography scheme. The originally combined templates are temporarily stored in temporary database. The fuzzy vault database stores the fuzzy helper data for fingerprint and commitment helper data for iris. As soon as the commitments or vaults are created, the templates are discarded. This also makes it hard to replay templates or Trojan to the system as the original biometrics was not stored.

### 2.4.5 EHR database

This database contains the text based records of patients. The portion of the EHR available to any request at any point in time depends on the privacy settings or rule presented by the system.

*2.4.6 Accountability database*

Authorised physicians can perform basically reading and writing to a patient EHR. When any of these defined operations are performed, the details of the task performed are documented by the system in hash format. For example, if a physician maliciously risked a patient's privacy by gaining malicious access to the confidential record without his or her consent, a trace is created in the accountability database. This ensures that patients are protected by the system.

*2.4.7 National healthcare server*

The national healthcare server in the framework of the developed system provides the only interface to the accountability phase of the system. This is to ensure that patients with necessary permission can request detail information about suspected breach to their EHR. The accountability system does not reside within a local hospital and the contents are protected using a secure hash algorithm.

*2.5 Unified modelling language (UML) analysis of the system*

The structure of the bio-cryptography key management model was identified using the unified modelling language (UML) analysis model. The model analysis led to the derivation of scenario based elements of the modelling elements, which are: the Use Case diagram and activity diagram of the system. The scenario is just a brief user story explaining who is using the system and what they are trying to accomplish. The UML Use Case diagram, flowchart and activity diagrams are all used to represent the bio-cryptography key management system structure.

Figure 3 shows the developed system use case analysis with the various actors, association, includes, extends and use cases associated with them depicted. This is a behaviour diagram that visually describes the functional requirements of the bio-cryptography key management model in EHR system and the relationships between actors and use cases. The three main actors are the patient, physician and the patient based escrow system for patient key management. There are four basic Includes-Use-Cases and four Extends-Use-Cases in the UML Use Case model. Figure 4 shows the overall business level activity diagram. This present the operation of the system starting from a patient swimlane where the cryptographic key to protect confidential records is generated to physician swimlane where refined EHR are accessed using the proposed privacy and security scheme. Figure 5 showed the flowchart of the bio-cryptography key management model in EHR system.

*2.6 Database model of the system*

Relational modelling technology was adopted for the design of the EHR system database. Microsoft SQL Server 2008 database engine was used to develop the database. The EHR relational model represents a description of some relvars and their attributes. In this design there are 20 relvars: tbleEHR, tblCBiodataC, tblRecType, tblPatientCStatusesCD, tblPatientLogin, tblPatientCSurgicalCD, tblPatientCBiodataCD, tblOperation, tblPatientCMedicalCD, tblHash, tblFV, tblEnscrowAgency, tblCStatusesC, tblBiocrypt, tblPatientCPsychiatricCD, tblPatientCPsychiatricCD, tblDoctorLogin, tblCSurgicalC, tblPsychiatricC, and tblCMedicalC. Each of the relvars is immediately followed by its tuples grouped into relations. The bold, underlined attributes are candidate keys. The non-bold, underlined attributes are foreign keys. The model relvars and their attributes are listed as follows:

i     tbleEHR (**eEHRID**, PatientNo, BiodataC, CMedicalC, PsychiatricMC, CSurgicalC, StatusesC)

ii    tblCBiodataC (**BiodataID**, Biodata, Basic, Confidential, Emergency)

iii   tblRecType (**RecID**, RecTypeNo, RecType)

iv    tblPatientCStatusesCD (**StatusesID**, PatientNo, [HIV/AIDS], [Blood group], Genotype, [Hepatitis B], [Hepatitis C] )

v     tblPatientLogin (**LoginID**, PatientNo, Password)

vi    tblPatientCSurgicalCD (**SurgeryID**, PatientNo, [Minor past sugeries], [Surgical implants], [Benigh Prostatic Hyperlasia] )

vii   tblPatientCBiodataCD (**PatientID**, PatientNo, LicenseNo, HospitalID, Firstname, Lastname, DOB, Gender, Religion, Nationality, MaritalStatus, Parity, Sexuality, PatientPic, Basic, Confidential, Emergency, Biometrics)

viii  tblOperation (**OperationID**, OperationNo, LicenseNo, Date, PatientNo, RecTypeNo)

ix    tblPatientCMedicalCD (**MedicalConditionID**, PatientNo, Hypertension, Diabetes, [Dyslipidemia/Hypercholerolemia], Arthritis, Arrhythmia, [Chronic kidney disease], Cancer, [Recurrent urinary tract infection], [Chronic obstructive pulmonary disease (COPD)], [Medical implant], Asthma, [Congestive heart failure], [Myocardial infarction angina], [Coronary artery disease], [Inflammatory bowel disease], [Parkinson disease], Epilepsy)

x     tblHash (**HashID**, OperationNo, Operation, OperationHash, LicenseNo)

xi    tblFV (**FVID**, LicenseNo, PatientNo, TempHex, FV, Status)

xii   tblEnscrowAgency (**EscrowID**, PatientNo, PatientKey, LicenseNo, CreatedNow, CreatedExpired, HospitalID, KeyStatus, EnrolledYet)

xiii  tblCStatusesC (**StatusesID**, Statuses, Basic, Confidential, Emergency)

xiv   tblBiocrypt (**BiocryptID**, HexaBio, BioKey, PatientNo, LicenseNo)

xv    tblPatientCPsychiatricCD (**PsychiatricID**, PatientNo, Autism, Mania, [Depressive illness], Schizophrenia)

xvi   tblDoctorBio (**DoctorID**, LicenseNo, HospitalID, Firstname, Lastname, Gender, DoctorPic, DOB, DoctorBiometrics)

xvii  tblDoctorLogin (**LoginID**, LicenseNo, Password)

xviii tblCSurgicalC (**CSurgicalCID**, CSurgicalC, Basic, Confidential, Emergency)

xix     tblPsychiatricC (**PsychiatricMCID**, PsychiatricMC, Basic, Confidential, Emergency)

xx     tblCMedicalC (**CMedicalCID**, CMedicalC, Basic, Confidential, Emergency).

**Figure 3**    Use case model of the developed bio-cryptography key management system

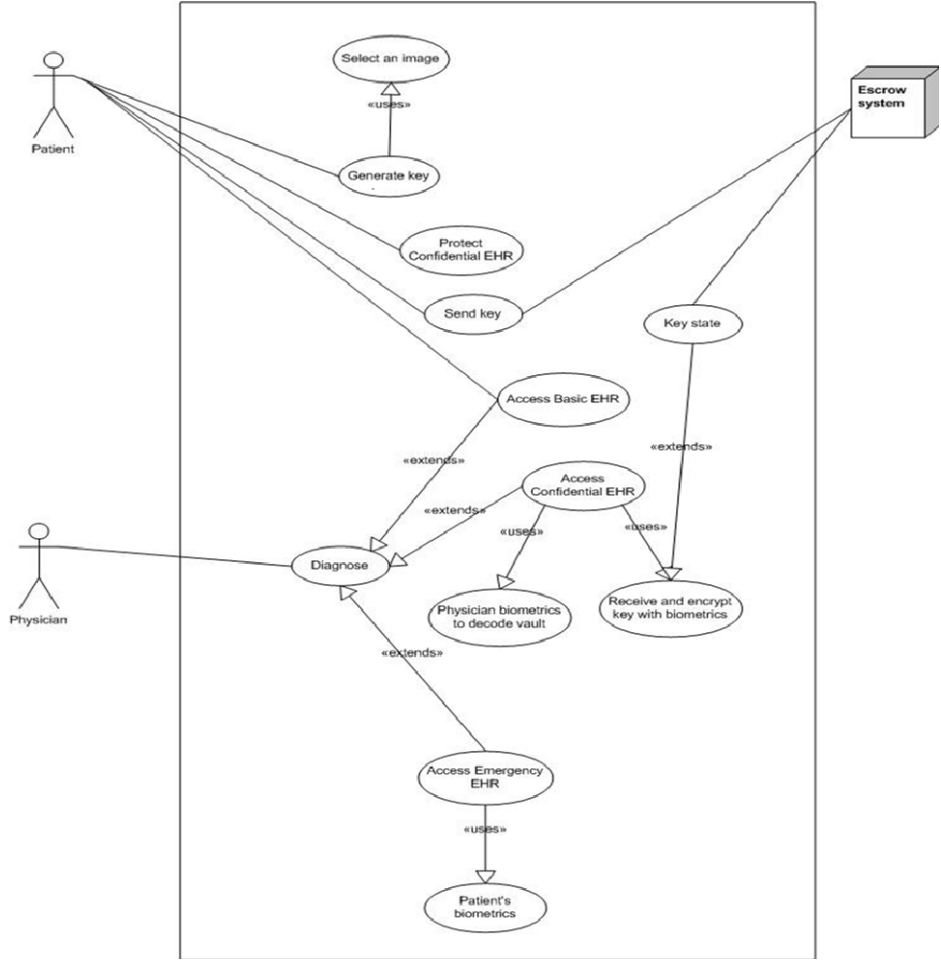

## 3 Findings and discussion

Privacy of patients was preserved by enforcing sharing rule in the EHR implementation so as to limit the number of shared or exposed data about the patients. Patients' right to decide who access their health data means the protection of their privacy. Such access will be made available for a particular period of time or session and could be revoked by the patients at their own discretion. The EHR view for basic records is shown in Figure 6; this section of the EHR does not require any special access besides the physician login. The emergency view is shown in Figure 7, this requires a patient biometrics for the records to be decrypted and displayed. Figure 8 shows the confidential records section of the designed her, which requires a decryption key from the data owner to be deciphered.

**Figure 4**  Business level activity diagram of the developed bio-cryptography system

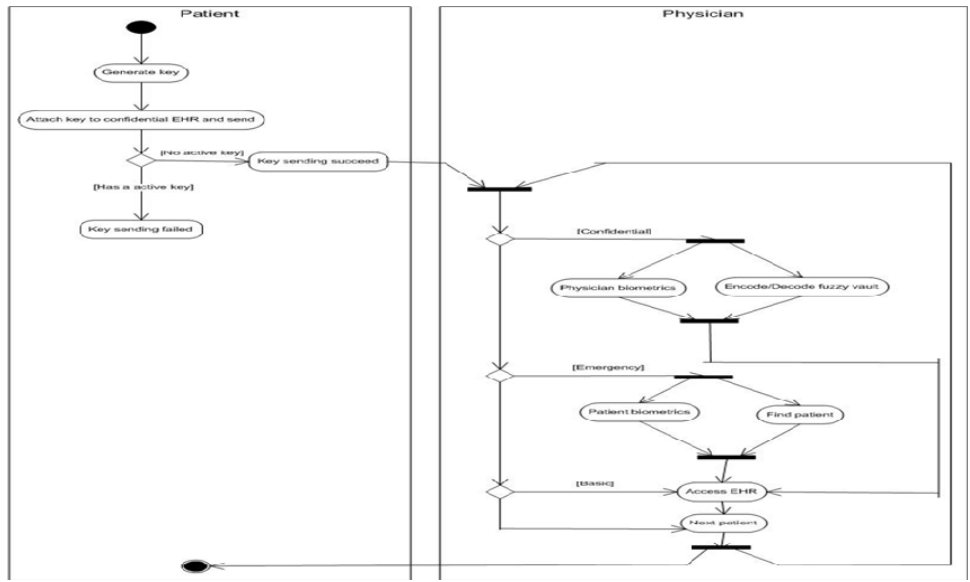

Four standard key states were implemented: key generation (via GLCM keys), key activation (via enrolled and active), key expiration (via enrolled/unenrolled and expired) and key archival (via expired). Whenever keys are acknowledged by the recipients, they are updated as 'ENROLLED'. A key can as well expire if it remained 'UNENROLLED' for the duration specified. A key remains 'ACTIVE' whether enrolled or not as long as the set duration by the patient has not lapsed.

When a receiving physician for a key is selected, the system enforces that the physician can only receive one active key per patients. In contrast, more than one physician can have separate active keys per patient as defined by the time specified. Each key generated via the modified GLCM method by the patient are stored in an escrowing agent database, the escrow agent act as an interface between a patient and his or her hospital. The unique timing feature is embedded into the proposed GLCM key generation algorithm in order to guarantee patients' privacy. This implies that, the patient can decide for themselves, the lifespan of their GLCM keys. However, the privacy of the receiving physicians is also protected as their *licenseno* are not available in decrypted form in the escrow system.

Figure 9 shows some of the sample keys with other attributes as stored in patients owned third party repository for keys. These keys cannot be used by this repository as it only identifies its clients who are the patients and ensures that patient's keys are updated or withdrawn at the patient's wishes. The use of escrow agents was enforced to solve the issue of discretional access control and poor delegation of confidential record keys. Also, patients do not have to install a personal key management system to keep records of their keys. For the purpose of demonstration, the keys stored in the escrow database are not encrypted.

**Figure 5** Flowchart of the developed bio-cryptography system (see online version for colours)

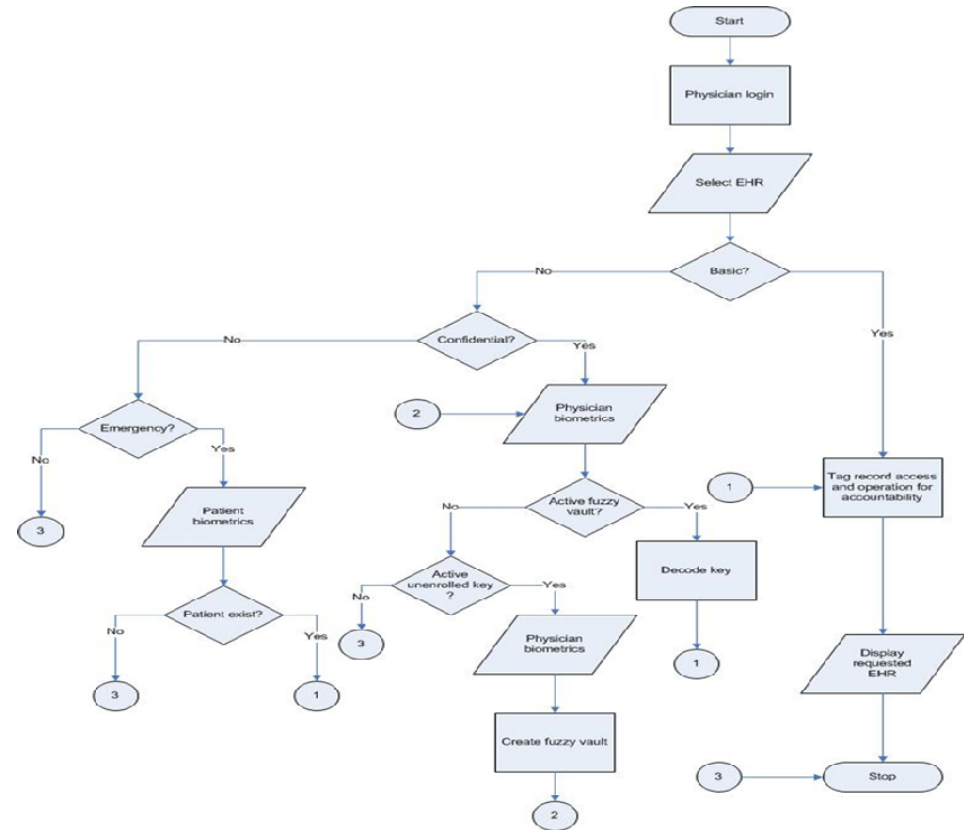

**Figure 6** Physician view of basic EHR (see online version for colours)

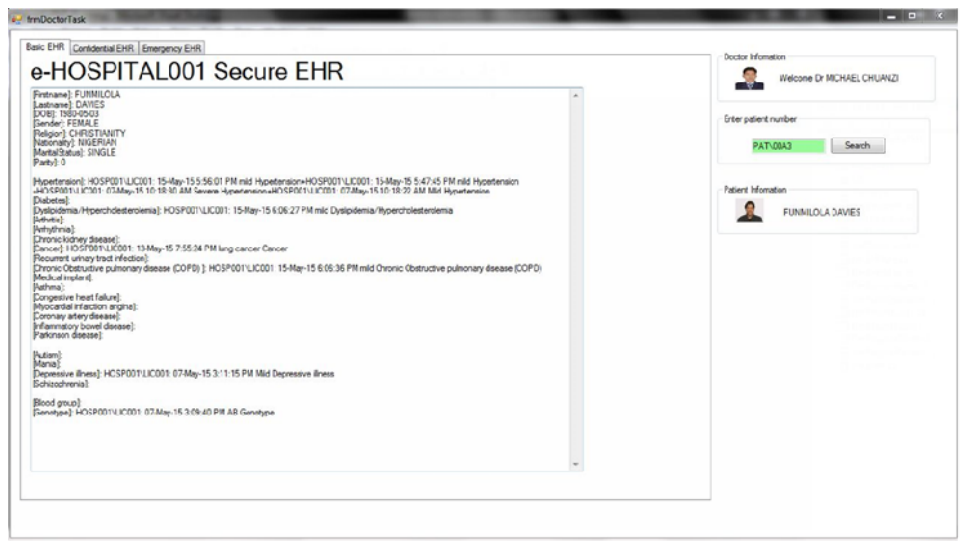

**Figure 7**   Physician view of emergency EHR (see online version for colours)

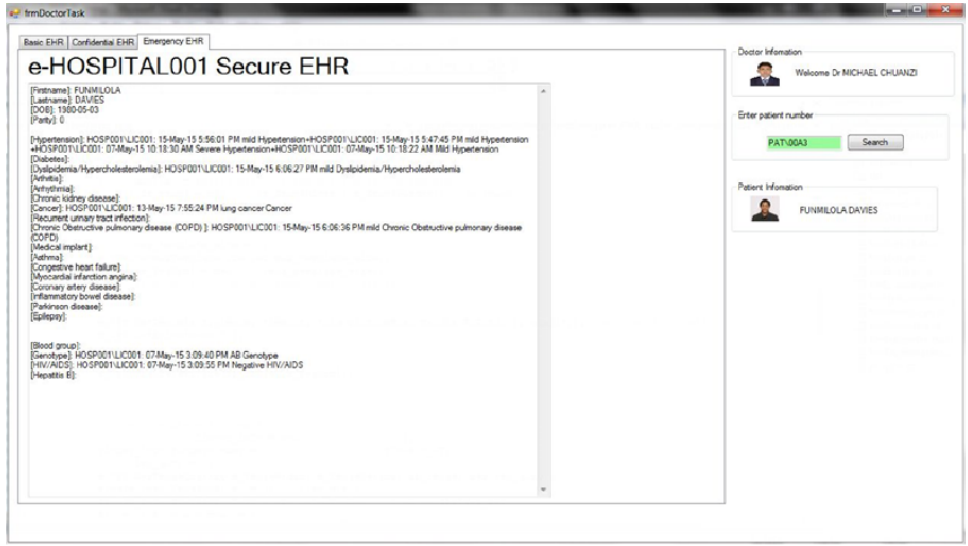

**Figure 8**   Patient view of confidential EHR (see online version for colours)

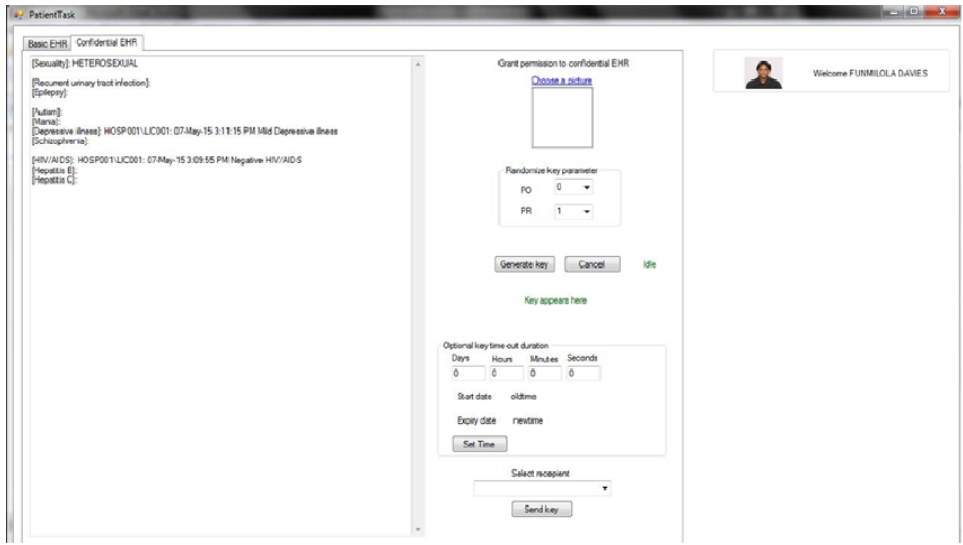

### 3.1 Implementation results of the modified fuzzy vault algorithm for key management

When a physician logs into the system, only basic EHR data as defined by the EHR rules embedded into the system are available. Other records categorised as confidential and emergency requires special access to decrypt. While confidential records require an active and enrolled key from the patient, emergency records require a biometrics of the patient to decrypt. If a patient has sent a key, such a key becomes available for enrolment so that it can be used by the doctor in the future to access confidential EHR as long as the

*Ensuring patients' privacy in a cryptographic-based-electronic health records* 17key is active. An example of a basic record derived from the EHR rule is shown in Figure 6.

**Figure 9** Patient's key information with escrow system (see online version for colours)

| EscrowID | PatientNo | PatientKey | LicenseNo | CreatedNow | CreatedExpired | HospitalID | KeyStatus | EnrolledYet |
|---|---|---|---|---|---|---|---|---|
| 7 | PAT\00A3 | 296435908232393 | 2e1340ed1d333... | 09-May-15 7:25... | 09-May-15 7:25... | HOSP001 | EXPIRED | UNENROLLED |
| 8 | PAT\00A3 | 327784155911915 | 2e1340ed1d333... | 11-May-15 3:42... | 11-May-15 3:45... | HOSP001 | EXPIRED | UNENROLLED |
| 9 | PAT\00A3 | 332695468255678 | 2e1340ed1d333... | 12-May-15 2:52... | 12-May-15 2:55... | HOSP001 | EXPIRED | UNENROLLED |
| 10 | PAT\00A3 | 184295079490945 | 2e1340ed1d333... | 12-May-15 7:11... | 12-May-15 7:14... | HOSP001 | EXPIRED | ENROLLED |
| 11 | PAT\00A3 | 266930262164517 | 2e1340ed1d333... | 12-May-15 7:20... | 12-May-15 7:25... | HOSP001 | EXPIRED | ENROLLED |
| 12 | PAT\00A3 | 214365746475311 | 76bb9b06171d5... | 12-May-15 8:09... | 12-May-15 8:29... | HOSP001 | EXPIRED | ENROLLED |
| 13 | PAT\00A3 | 374719803420474 | 76bb9b06171d5... | 13-May-15 4:44... | 13-May-15 4:54... | HOSP001 | EXPIRED | ENROLLED |
| 14 | PAT\00A3 | 784969352837134 | 76bb9b06171d5... | 13-May-15 5:13... | 13-May-15 5:23... | HOSP001 | EXPIRED | ENROLLED |
| 15 | PAT\00A3 | 161717116762518 | 76bb9b06171d5... | 13-May-15 5:29... | 13-May-15 5:44... | HOSP001 | EXPIRED | ENROLLED |
| 16 | PAT\00A3 | 79810909702947 | 76bb9b06171d5... | 13-May-15 6:09... | 13-May-15 6:16... | HOSP001 | EXPIRED | ENROLLED |
| 17 | PAT\00A3 | 186786025960424 | 76bb9b06171d5... | 14-May-15 7:04... | 14-May-15 7:34... | HOSP001 | EXPIRED | ENROLLED |
| 18 | PAT\00A3 | 2476777558358 | 2e1340ed1d333... | 14-May-15 7:04... | 14-May-15 7:34... | HOSP001 | EXPIRED | ENROLLED |
| 19 | PAT\00A2 | 123695483253836 | 76bb9b06171d5... | 14-May-15 7:11... | 14-May-15 7:41... | HOSP001 | EXPIRED | ENROLLED |
| 20 | PAT\00A2 | 123695483253836 | 2e1340ed1d333... | 14-May-15 7:11... | 14-May-15 7:41... | HOSP001 | EXPIRED | ENROLLED |
| 21 | PAT\00A4 | 511456510545262 | 76bb9b06171d5... | 14-May-15 7:11... | 14-May-15 7:45... | HOSP001 | EXPIRED | ENROLLED |
| 22 | PAT\00A4 | 511456510545262 | 2e1340ed1d333... | 14-May-15 7:11... | 14-May-15 7:45... | HOSP001 | EXPIRED | ENROLLED |
| 23 | PAT\00A2 | 195319872134782 | 76bb9b06171d5... | 14-May-15 10:2... | 14-May-15 11:2... | HOSP001 | EXPIRED | ENROLLED |
| 24 | PAT\00A4 | 34034747935543 | 76bb9b06171d5... | 14-May-15 10:2... | 14-May-15 10:4... | HOSP001 | EXPIRED | UNENROLLED |
| 25 | PAT\00A2 | 198151730307531 | 2e1340ed1d333... | 14-May-15 11:1... | 14-May-15 11:1... | HOSP001 | EXPIRED | UNENROLLED |
| 26 | PAT\00A3 | 452453419640695 | 76bb9b06171d5... | 14-May-15 11:1... | 14-May-15 11:1... | HOSP001 | EXPIRED | UNENROLLED |
| 27 | PAT\00A3 | 786598942519134 | 76bb9b06171d5... | 14-May-15 11:2... | 14-May-15 11:2... | HOSP001 | EXPIRED | UNENROLLED |
| 28 | PAT\00A3 | 786598942519134 | 2e1340ed1d333... | 14-May-15 11:2... | 14-May-15 11:2... | HOSP001 | EXPIRED | UNENROLLED |
| 29 | PAT\00A2 | 874534806048759 | 2e1340ed1d333... | 14-May-15 11:2... | 14-May-15 11:2... | HOSP001 | EXPIRED | UNENROLLED |
| 30 | PAT\00A4 | 102279553959626 | 76bb9b06171d5... | 14-May-15 11:2... | 14-May-15 11:2... | HOSP001 | EXPIRED | UNENROLLED |

Physicians' biometrics is required to enrol a key and fuzzy vault helper data are consequently created. Although, the nature of fingerprint biometrics high false acceptance (FAR) could pose a big challenge when matching a vault, in order to avoid this, maliciously engineered enrolment is prevented. This EHR system design ensures that a doctor cannot use another person's biometrics to create a vault or enrol a key. This implies that a malicious insider who gained access to a doctor's system login detail would not be able to access patients' confidential keys with his or her biometrics. Figure 10 presents a malicious insider who has gained access to the login ID and password of a doctor but who is unable to enrol or use a key in order to fraudulently gained access to the genuine doctor's and patient's confidential records. If the right biometrics was provided, biometric key enrolment becomes successful and available for use as shown in Figure 11.

A successful match creates a fuzzy vault whose public helper data is a combination of the encrypted biometrics, polynomial coefficients and some random numbers. Segment of helper data is shown in Figure 12, which represents a section of the fuzzy vault database generated by the developed system algorithm. This model allows a single user to have multiple vault status, just as an encryption key could expire, a fuzzy vault helper data representing a particular key could also be invalid over time. The status of a fuzzy vault key is determined by the time set by the patient. Whenever physicians successfully enrolled keys with their biometrics, corresponding public fuzzy vault (FV) helper data are created. The helper is public in the sense that both the biometrics and the key it encodes remained open or visible in the database as helper data.

**Figure 10** Malicious insider who has gained access to the login ID and password of a doctor (see online version for colours)

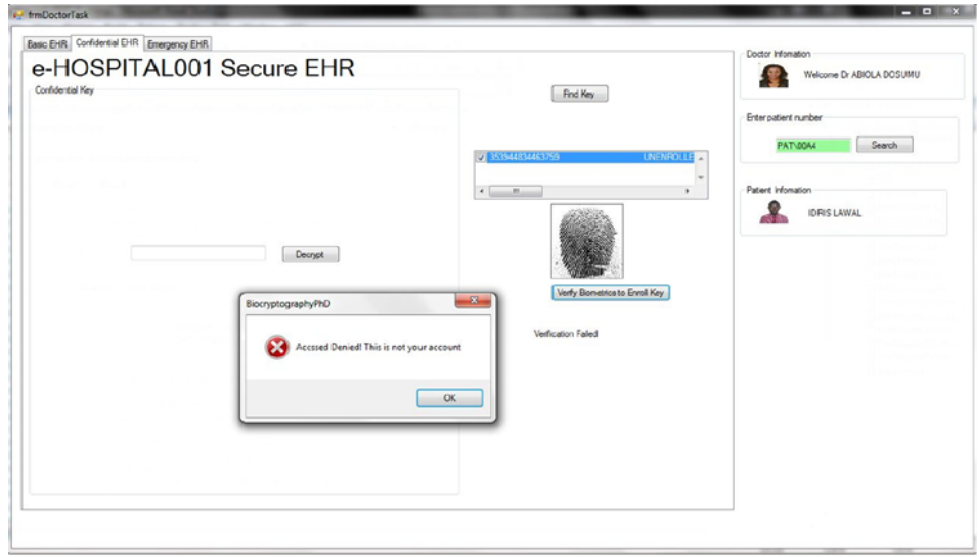

**Figure 11** Successful bio-cryptography key enrolments (see online version for colours)

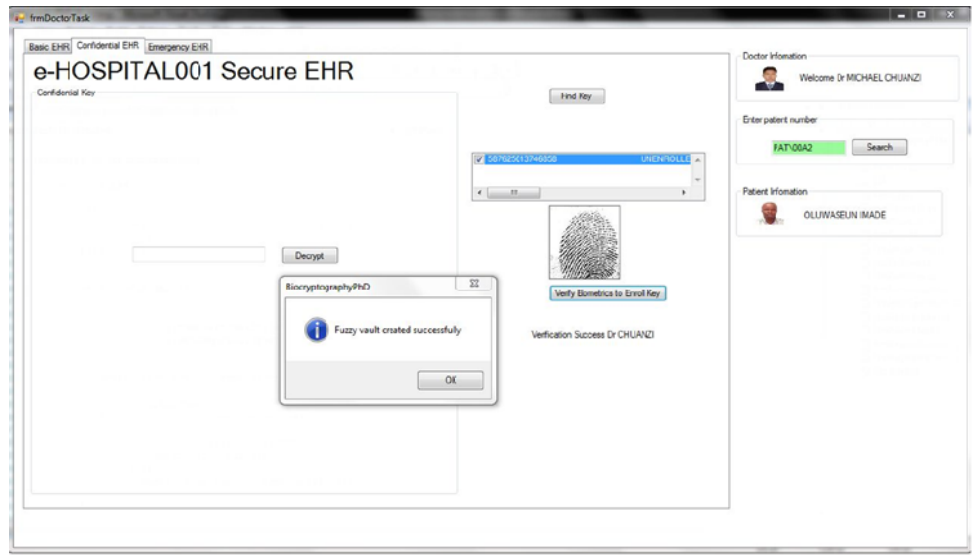

### 3.2 Secure hash monitoring

This work is limited to observe read and write operations. All malicious operations performed on patients' EHR are securely monitored using SHA-256 of the operation performed. Since SHA-256 hash is difficult to reverse engineered, it becomes impossible for malicious inside attackers to undo their operations. The ID of the suspect as well as the patient whose records are tampered with, were both stored as hash. This reduces the volume of logs created unlike in previous similar works, where logs are usually voluminous and requiring massive storage requirements. This work tags each hash operations together by using similar set of numbers {1,2,3} for operation and record type scrambling, this is to make interpretation difficult for anyone who maliciously gained access to the accountability database. Figure 13 shows the primary accountability. Log volumes are also lesser because operations are identified via tagging. The protection of issuer's privacy is also an improvement over the work of Mashima and Ahamad (2012) and accountability is also not limited to emergency which also represents another improvement over Sun et al. (2011).

**Figure 12** Segment of fuzzy vault helper data (see online version for colours)

| FVID | LicenseNo | PatientNo | Tem... | FV | Status |
|---|---|---|---|---|---|
| 6 | 2e1340ed1d333... | fcc27f1614da0c4c93a103a46746cd51642... | c3117... | c3117e078afce0f135ee9349bb603840c289d1799501a8c951c2e8dcb0abfe5... | EXPIRED |
| 8 | 76bb9b06171d5... | fcc27f1614da0c4c93a103a46746cd51642... | 5cfad... | 5cfadc5c3d661bb821d23843a4806a77b610965847eb55dd0d9c0171c1e11c... | EXPIRED |
| 9 | 76bb9b06171d5... | fcc27f1614da0c4c93a103a46746cd51642... | e1b16... | e1b1653c5366ea02b5431daf4829cb308b09c043c681980728941725a9bb87... | EXPIRED |
| 10 | 76bb9b06171d5... | fcc27f1614da0c4c93a103a46746cd51642... | 8817a... | 8817a391b4891b77f38b8c2a9a7dfe9070264377d5b24f246190c6036484d52... | EXPIRED |
| 11 | 76bb9b06171d5... | fcc27f1614da0c4c93a103a46746cd51642... | 4902d... | 4902d2ac5698d66956485e865b307e832c4ab7c85d35f8640c7b32e83d5cae... | EXPIRED |
| 12 | 76bb9b06171d5... | fcc27f1614da0c4c93a103a46746cd51642... | 10d1b... | 10d1b694c1e27397b94364f60d5457610dd6149dfb1968e7dd23a6f5ac0be6... | EXPIRED |
| 13 | 76bb9b06171d5... | fcc27f1614da0c4c93a103a46746cd51642... | dbe5f... | dbe5f7339da044e346787e166cb612e51d3b8b18820e844679c05121c3722fa... | EXPIRED |
| 14 | 76bb9b06171d5... | 80c2312e9ebab657bca5418108521e4b0d... | d5da6... | d5da6c716817f0be5b03f01566b6d743d447832b581f143329f5b990b24227f... | EXPIRED |
| 15 | 76bb9b06171d5... | d121400ac1f23ced2c43cbd1c55425fb27... | 38030... | 38030b5117cc5f391534177ea30b668fbd24ab4bce3228a5b7f407772d242f5... | EXPIRED |
| 16 | 2e1340ed1d333... | fcc27f1614da0c4c93a103a46746cd51642... | 15832... | 1583233d8c915d83457d3832a5f9ccced3022bffe8515c98f0811b7912b7ec9f... | EXPIRED |
| 17 | 2e1340ed1d333... | 80c2312e9ebab657bca5418108521e4b0d... | 51d0c... | 51d0c805dbb001c9d09951216f708a4a36158d936fae7fe0217a8308b07aa3a... | EXPIRED |
| 18 | 2e1340ed1d333... | d121400ac1f23ced2c43cbd1c55425fb27... | 948e7... | 948e74b673bd85eac7539774bba5de404e99910cd442e4a4ee63309cb7e0d2... | EXPIRED |
| 19 | 76bb9b06171d5... | 80c2312e9ebab657bca5418108521e4b0d... | 31a54... | 31a543dfde62fb8b9f07ba2f4b1e81a1b0619e63c2e124ea66e8965d7921d81... | EXPIRED |
| 20 | 2e1340ed1d333... | d121400ac1f23ced2c43cbd1c55425fb27... | 03754... | 03754dd65365bf86ace9713015c52684b8425a58cffa7c84d7bd4ae78dcecaa... | EXPIRED |
| 21 | 2e1340ed1d333... | 80c2312e9ebab657bca5418108521e4b0d... | 0d295... | 0d29512667e7f31310070715e082ee62f2f6d4c302c3b0a2c9e015f3dc6b3c13... | EXPIRED |
| 22 | 76bb9b06171d5... | 80c2312e9ebab657bca5418108521e4b0d... | 122b1... | 122b18794fd1a92ff0d0ac5a2ce0240fb5ee1269d33f4c212f953191fa4865caf... | EXPIRED |
| 23 | 76bb9b06171d5... | fcc27f1614da0c4c93a103a46746cd51642... | 0a7ac... | 0a7ac3da195b25d02ca173deeaa403bf609e92aba0b754122d5775559c4d83... | ACTIVE |
| 24 | 2e1340ed1d333... | fcc27f1614da0c4c93a103a46746cd51642... | 9122b... | 9122b9d87eacfc09aeecce1131576a6c1fcd4492acfce43fc3ab0774d63ff813... | ACTIVE |
| 25 | 76bb9b06171d5... | 80c2312e9ebab657bca5418108521e4b0d... | 7db10... | 7db10412efe86625fb2c0566c694b1113c4dda5f7912cd149e2e67131afe609... | ACTIVE |

| 1 of 22 | Cell is Read Only.

An example of a write operation is shown in Figure 14 and emergency access is shown in Figure 15, requiring patient biometrics. A write operation occurs when during e-consultation or physical consultation when a physician added new results of diagnosis to a patient record. Besides monitoring, every single diagnosis added to a patient EHR bears the signature of the updater. Also, a timestamp, which is clearly readable, is also added to the new modification made to the records.

The system design highly disfavours identifying a patient with a wrong biometrics during emergency. This is because having access to not just timely information is important when dealing with lives of individuals, accurate information is as well important so that patients can be diagnosed correctly during emergency. The effect of correct identification involves trading a high FAR for moderate false rejection. The FRR is moderated by ensuring that more that 80% of an enrolment minute were used for verification. The implication of this approach was recorded in the system performance evaluation.

20      *A. Omotosho et al.*

**Figure 13**  Primary accountability database (see online version for colours)

**Figure 14**  A write operation into an EHR (see online version for colours)

### 3.3  System evaluation

Live biometrics samples of consisting of 200 fingerprints and 200 irises were captured and used to test the efficiency of the fuzzy vault and fuzzy vault algorithms. Evaluation metrics: FAR, FRR and time complexity of the algorithms execution were recorded accordingly.

**Figure 15** Biometrics accessing of emergency data (see online version for colours)

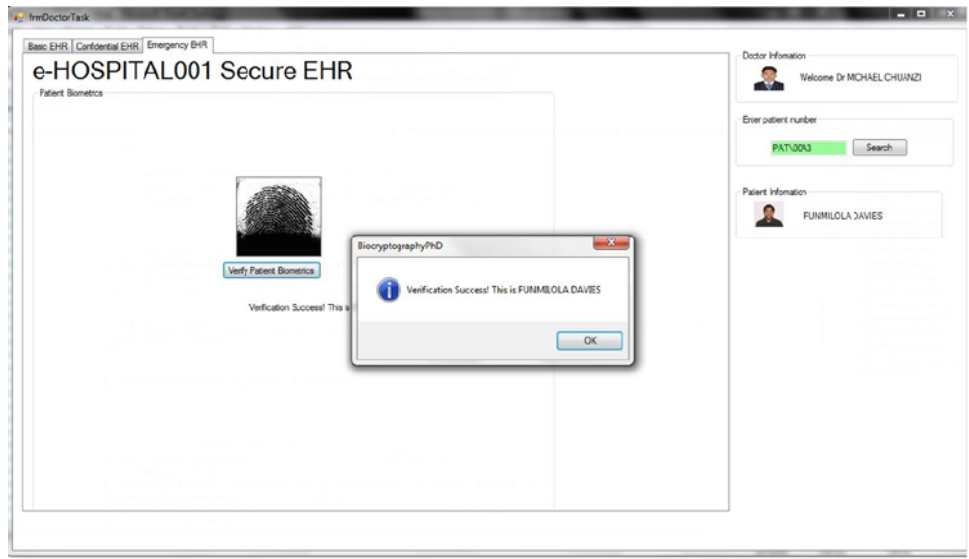

For most biometric systems, FRR ranges from 0.1% to 20%, meaning that a legitimate user will be rejected from one out of 1000 times to one out of 5 times on average. FAR ranges from one in 100, for 'low security applications', to one in 10,000,000, for 'very high security' applications. The FAR of the fingerprint fuzzy vault is 0% while the FRR is 2%. The system FAR of 0% which is extremely low for fingerprint identification and this implies keys will not likely be available to malicious users by trial and error or brute force approach.

Most of the work in literature emphasises accuracy and not on the time cost of encrypting and decrypting a vault especially when using the slow error correcting codes to decipher a vault or commitment. The total encryption time for the fuzzy vault was 24.914 s with average of 0.125 ms per vault. The total decryption time was 459.284 s with an average vault decryption time of 2.296 s per helper data. Figure 16 shows the execution time of the fingerprint fuzzy vault algorithm with the four rejected biometrics. A segment of the experimentation data are illustrated in Figure 17, where 1 indicates a correct acceptance while 0 represent a false rejection.

There were higher discrepancies between iris code encoding and decoding time. Hamming distance threshold of 0.3 is used in order to identify iris codes that are sufficiently close. The iris fuzzy commitment has a total encoding time of 7.789 s and an average encoding time of 0.0389 ms which represent a better performance compare to the fingerprint fuzzy vault values (total: 24.914 s and average: 0.125 ms). However, the total decoding time is 610 s with average decoding time of 3.051 s; these values were higher than the ones obtained from fuzzy vault evaluation (total: 459.284 s total and average: 2.296 s). Even though, iris fuzzy commitment appeared to be faster when encoding the 200 iris codes, overhead of 150.716 s for just 200 irises could become a challenge when the iris code dataset becomes larger. This unanticipated higher decoding time contributed to one of the reasons fuzzy vault is preferred in the developed bio-cryptography key management in EHR.

**Figure 16** Fuzzy vault execution, acceptance and rejection graph (see online version for colours)

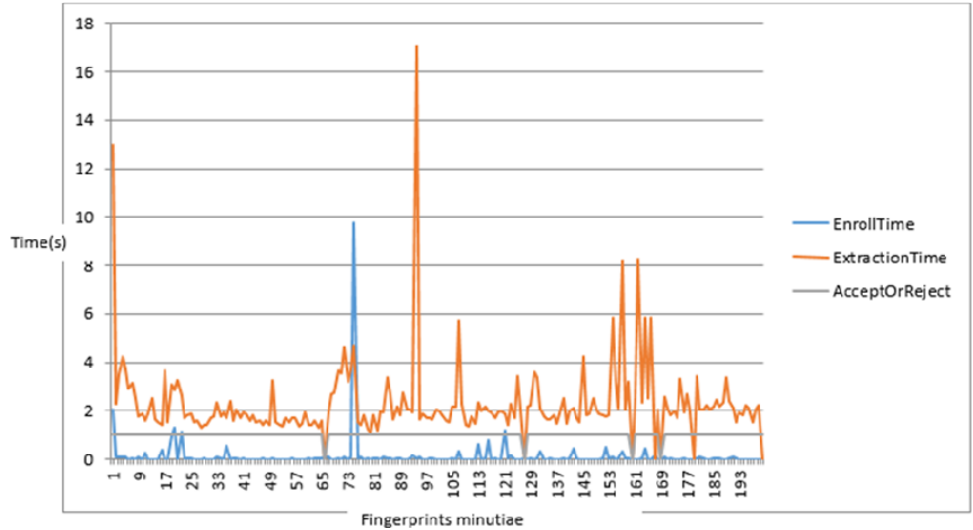

**Figure 17** A segment of the fuzzy vault evaluation database (see online version for colours)

Figure 18 shows the total encoding and decoding time for the commitment together with the acceptance and rejection rate. Similar to fuzzy vault, a FAR of zero is recorded. The fuzzy commitment FRR is 10% which is higher than the best known iris fuzzy commitment by Hao et al. (2006) where similar commitment scheme was used with Daugman algorithm for iris codes. The result is however better than FRR of 50.0% and a FAR of 7.0% for iris biometrics obtained later by Maiorana and Ercole (2007) among others as displayed in Table 2. One major challenge is due to the difficultly in constructing high quality iris codes of shorter bit streams.

**Figure 18** Fuzzy commitment execution, acceptance and rejection graph (see online version for colours)

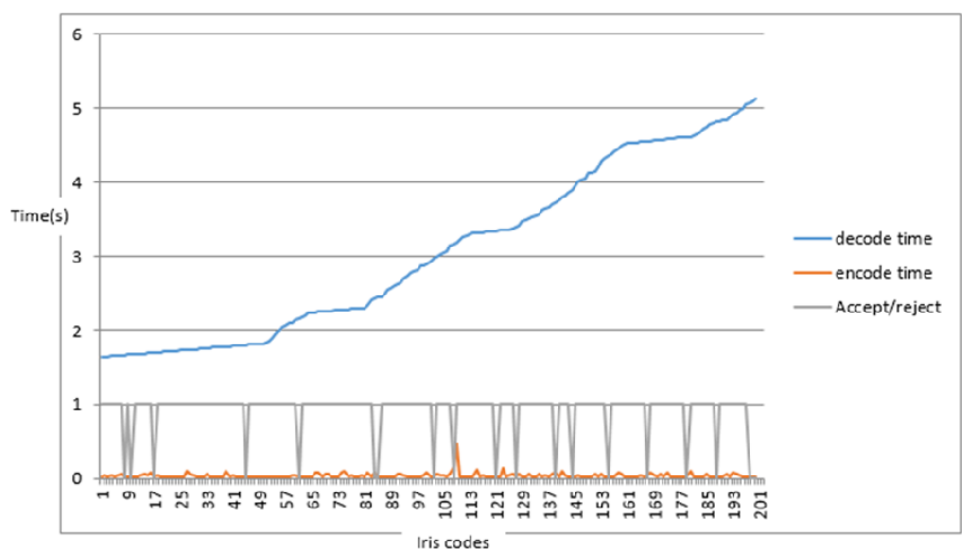

The BuIris contains $640 \times 480$ pixel iris image dimension and in .bmp format. All the templates used in this work were converted to conform with the standard ISO 19794-2:2005 for fingerprints and ISO/IEC 19794-6:2005 for iris data. These formats are generic in that they may be applied and used in a wide range of application areas where automated fingerprints minutiae and iris code are to be captured.

## 4 Implications and limitations

Technologies have metamorphosed the role of patients from the conventional passive receiver of healthcare services into a more active role in which they have more understanding of their medical records and are empowered with the ability to make certain choices and be involved in decisions making. This study has provided a support for active patients' participation in maintaining privacy and security in EHR systems that relied solely on cryptography for achieving them (Benaloh et al., 2009; Sun et al., 2011; Li et al., 2012; Dharanya et al., 2013). The sophistication of these systems depends on the secrecy of encryption keys and this is one of the issues addressed in this study. The proposed scheme in this work can increase patients' trust of cryptographic-based-EHR.

Encryption keys methods used in this study will enable patients to generate keys that can be used by anyone they want to give confidential access for specified durations. This satisfies the purpose of patients' privacy to exercise control on the limit to which access to their confidential information is available.

**Table 2**     Experimental results of fuzzy vault and fuzzy commitment

| Authors | Applied technique | Modality | FRR/FAR (%) | Encryption/ decryption time | Database | Datasets | Remark |
|---|---|---|---|---|---|---|---|
| Hao et al. (2006) | Fuzzy commitment | Iris | 0.42/0.0 | – | Proprietary | 70 | Small test set with 140bits key |
| Bringer et al. (2007) | Fuzzy commitment | Iris | 5.62/0.0 | – | ICE | 244 | Short key with 40 bits key |
| Clancy et al. (2003) | Fuzzy vault | Fingerprints | 20–30/0.0 | – | – | | 224bit pre-alignment, >1 enrol sample |
| Nandakumar et al. (2007) | Fuzzy vault | Fingerprints | 4.0/0.004 | – | FVC2002-DB2 | 110 | >1enrol sample |
| Maiorana and Ercole (2007) | Fuzzy commitment | Iris | 50.0/7.0 | – | – | | – |
| This research | Fuzzy vault | Fingerprint | 2.0/0.0 | 0.125 ms/ 2.296 s | BuIris | 200 | 120bits key; 1 enrol samples |
| This research | Fuzzy commitment | Iris | 10.0/0.0 | 0.0389 ms/ 3.051 s | BuIris | 200 | 120bits key 1 enrol samples |

The essence of the predefined confidential attributes of EHR saves individual practitioner from the stress of having to repeatedly decide what a patient can protect at any point in time. In the design of the EHR, some of the attributes selected by the practitioners were overlapped, this redundancy ensures that some attributes of EHR appears in more than one sections (basic, confidential and emergency) covered in this study. The benefit of this to practitioners is that, every record that needs to be available for patients' general treatments are available. For example, attributes like genotype is always in any record while HIV status is only available in confidential and emergency.

This study presents one of the few deployments and empirical analysis of the complex fuzzy vault and fuzzy commitment schemes in an application. Evaluation results are very positive in terms of the access time and accuracy especially for the fingerprints fuzzy vault. This means that in a time and precision dependent application like the EHR, bio-cryptography can be used to secure keys and patients privacy without much performance degradation. In addition, other researchers can deploy similar or improved methods in several other real life applications based on the results obtained in this study.

The evaluation of the time-dependent GLCM key algorithm is not carried out in even though, keys were generated from small region of the image to lower the computational overhead of processing a full image. It is important that future work considers the experimentation of the time cost that it will take patients to generate their confidential keys.

## 5 Conclusion

Privacy and security are two important factors that must be considered when developing a patient centric EHR. Existing works have focused largely on security and less on how patient mastermind privacy control with the help of the healthcare providers. Also, outside threats were of more concern, but more often than expected insiders who have legitimate access to EHR are often overlooked. This research implements cryptographic techniques with biometrics despite its key security challenges. This research contributed to the body of knowledge in healthcare and self-care through the development of a bio-cryptography key management scheme for ensuring the privacy of patients attributes in a cryptographic oriented EHR. The result of the system evaluation shows significant improvements as the biometrics FAR were greatly reduced which lessens the likelihood of imposters gaining successful access to patients protected health information. This result also justifies the feasibility of implementing fuzzy key binding scheme in real applications, especially fuzzy vault, which demonstrated a high performance from evaluation.

## References

Appari, A. and Johnson, M.E. (2010) 'Information security and privacy in healthcare: current state of research', *International Journal of Internet and Enterprise Management*, Vol. 6, No. 4, pp.279–314.

Benaloh, J., Chase, M., Horvitz, E. and Lauter, K. (2009) 'Patient controlled encryption: ensuring privacy of electronic medical records', *Proceedings of the 2009 ACM Workshop on Cloud Computing Security*, New York, USA.

Bhartiya, S. and Mehrotra, D. (2015) 'An access control framework for secured sharing of electronic health records using hierarchy similarity analyser', *Int. J. Electronic Healthcare*, Vol. 8, Nos. 2–4. pp.121–141.

Bringer, J., Chabanne, H., Cohen, G., Kindarji, B. and Zemor, G. (2007) 'Optimal iris fuzzy sketches', *Proceedings of 2007 First IEEE International Conference on Biometrics: Theory, Applications and Systems*, Crystal City, Virginia.

Brumen, B., Heričko, M., Sevčnikar, A., Završnik, J. and Hölbl, M. (2013) 'Outsourcing medical data analyses: can technology overcome legal, privacy and confidentiality issues?', *J. Med. Internet Res. Journal of Medical Internet Research*, Vol. 15, No. 12, pp.283.

Carroll, R. (2016) *Aspen Valley Hospital Accused of Patient-Privacy Breach*, http://www.aspentimes.com/news/22463520-113/aspen-valley-hospital-accused- of-patient-privacy-breach (Accessed 18 July, 2016).

Clancy, T.C., Kiyavash, N. and Lin, D.J. (2003) 'Secure smartcard-based fingerprint authentication', *Proceedings of ACM SIGMM 2003 Multimedia, Biometrics Methods and Applications Workshop*, Berkeley, California.

Collins, K. and Robertson, J. (2014) *Bloomberg Visual Data: Data Breaches in the U.S.*, http://www.bloomberg.com/infographics/2014-08-21/top-data- breaches.html (Accessed 30 August, 2014).

Computer Economics (2007) 'Computer economics report', *Computer Economics*, Vol. 29, No. 4, pp.1–5.

CTI (Center for Technology Innovation at Brookings) (2016) *Hackers, Phishers, and Disappearing Thumb Drives: Lessons Learned From Major Health Care Data Breaches*, http://www.brookings.edu/~/media/research/files/papers/2016/04/28-patient-privacy-yaraghi/patient-privacy504v3.pdf (Accessed 18 July, 2016).

Das, R. (2012) *Bio-Cryptography: Increasing the Protection of Biometric Templates*, http://www.biometricnews.net/article3.pdf (Accessed 30 August, 2014).

Demiris, G. (2004) 'Electronic home healthcare: concepts and challenges', *Int. J. Electronic Healthcare*, Vol. 1, No. 1, pp.4–16.

Dharanya, S., Indira, P.D. and Blessy, S. (2013) 'Achieving secure personal health records using multiple-Authority attribute based encryption', *International Journal of Research in Engineering and Advanced Technology*, Vol. 1, No. 1, pp.1–5.

Dinev, T., Albano, V., Xu, H., D'Atri, A. and Hart, P. (2016) 'Individuals' attitudes towards electronic health records: a privacy calculus perspective', *Advances in Healthcare Informatics and Analytics Annals of Information Systems*, pp.19–50.

Escarfullet, K., Moore, C., Tucker, S. and Wei, J. (2012) 'An object-oriented mobile health system with usability features', *International Journal of Electronic Healthcare*, Vol. 7, No. 1, pp.53–67.

Gordon, L.A., Loeb, M.P., Lucyshyn, W. and Richardson, R. (2006) *2006 CSI/FBI Computer Crime and Security Survey*, Computer Security Institute, pp.4–9.

Grunwel, D. and Sahama, T. (2016) 'Delegation of access in an information accountability framework for eHealth', *Proceedings of the Australasian Computer Science Week Multiconference*, Canberra, Australia.

Hao, F., Anderson, R. and Daugman, J. (2006) 'Combining cryptography with biometrics effectively', *IEEE Transactions on Computers*, Vol. 55, No. 9, pp.1081–1088.

Harris, K.D. (2016) *California Data Breach Report 2012-2015*, https://oag.ca.gov/sites/all/files/agweb/pdfs/dbr/2016-data-breach-report.pdf (Accessed 20 July, 2016).

Hewapathiranae, R. (2011) 'Electronic health records: convenience vs potential security vulnerability', *Sri Lanka Journal of Bio-Medical Informatics*, Vol. 2, No. 2, pp.38–40.

Hewitt, C. (2013) 'For privacy and security, use public keys everywhere', *Communications of the ACM*, Vol. 56, No. 9, p.8.

Hillestad, R., Bigelow, J., Bower, A., Girosi, F., Meili, R., Scoville, R. and Taylor, R. (2005) 'Can electronic medical record systems transform health care? Potential health benefits, savings and costs', *Health Affairs*, Vol. 24, No. 5, pp.1103–1117.

HIMSS (Healthcare Information and Management Systems Society) (2003) *EHR Definition, Attributes and Essential Requirements*, http://www.himss.org/content/files/ EHRAttributes.pdf (Accessed 8 August, 2014).

ICIT (Institute for Critical Infrastructure Technology) (2016) *Hacking Healthcare IT in 2016: Lessons the Healthcare Industry Can Learn From the OPM Breach*, http://icitech.org/wp-content/uploads/2016/01/ICIT-Brief-Hacking-Healthcare-IT-in- (2016).pdf (Accessed 18 July, 2016).

IMIA (International Medical Informatics Association) (2000) 'Recommendations of the international medical informatics association (IMIA) 'on education in health and medical informatics', *Methods Archive*, Vol. 39, No. 3, pp.267–277.

Juels, A. and Sudan, M. (2002) 'A fuzzy vault scheme', *Proceedings of IEEE International Symposium on Information Theory*, Lausanne, Switzerland.

Juels, A. and Wattenberg, M. (1999) 'A fuzzy commitment scheme', *Proceedings of the 6th ACM Conference on Computer and Communications Security*, Kent Ridge Digital Labs, Singapore.

Li, M., Yu, S., Zheng, Y., Ren, K. and Lou, W. (2013) 'Scalable and secure sharing of personal health records in cloud computing using attribute-based encryption', *IEEE Transactions on Parallel and Distributed Systems*, Vol. 24, No. 1, pp.131–143.

Maiorana, E. and Ercole, C. (2007) 'Secure biometric authentication system architecture using error correcting codes and distributed cryptograph', *Rapport 2007 Annual Meeting of the Association Group National Telecommunications and Information Technologies (GTTI'07)*, Roma, Italy, pp.1–12.

Mashima, D. and Ahamad, M. (2012) 'Enhancing accountability of electronic health record usage via patient-centric monitoring', *Proceedings of the 2nd ACM SIGHIT International Health Informatics Symposium (IHI '12)*, Miami, Florida.

Mercuri, R.T. (2004) 'The HIPAA-potamus in health care data security', *Communications of the ACM*, Vol. 47, No. 7, p.25.

Nandakumar, K., Jain, A.K. and Pankanti, S. (2007) 'Fingerprint-based fuzzy vault: implementation and performance', *IEEE Transactions on Information Forensics and Security*, Vol. 2, No. 4, pp.744–757.

Omotosho, A. and Emuoyibofarhe, J. (2014) 'A criticism of the current security, privacy and accountability issues in electronic health records', *International Journal of Applied Information Systems*, Vol. 7, No. 8, pp.11–18.

Omotosho, A. and Emuoyibofarhe, J. (2015) 'Private key management scheme using image features', *Journal of Applied Security Research*, Vol. 10, No. 4, pp.543–557.

Rathgeb, C. and Uhl, A. (2011) 'A survey on biometric cryptosystems and cancelable biometrics', *EURASIP Journal on Information Security*, Vol. 20, No. 1, pp.3–25.

Robertson, J. (2014) *China's Hack of 4.5 Million, U.S. Medical Records? This Chart Will Make You Sick*, http://www.bloomberg.com/news/2014-08-21/china-s- hack-of-4-5-million-u-s-medical-records-this-chart-will-make-you-sick.html (Accessed 30 August, 2014).

Sanzgiri, A. and Dasgupta, D. (2016) 'Classification of insider threat detection techniques', *Proceedings of the 11th Annual Cyber and Information Security Research Conference*, Oak Ridge, Tennessee.

Scheirer, W., Bishop, B. and Boult, T. (2010) 'Beyond PKI: the biocryptographic key infrastructure', *IEEE International Workshop on Information Forensics and Security (WIFS)*, Seattle, Washington, pp.1–6.

Srinivasan, S. (2016) 'Compromises in healthcare privacy due to data breaches', *European Scientific Journal*, Vol. 12, No. 10, pp.91–98.

Stavroulakis, P. and Stamp, M. (2010) *Handbook of Information and Communication Security*, Springer, Heidelberg.

Sun, J., Zhu, X., Zhang, C. and Fang, Y. (2011) 'HCPP: cryptography based secure EHR system for patient privacy and emergency healthcare', *IEEE 31st International Conference on Distributed Computing Systems*, Minneapolis, Minnesota, pp.373–382.

Tiwari, B. and Kumar, A. (2015) 'Role-based access control through on-demand classification of electronic health record', *Int. J. Electronic Healthcare*, Vol. 8, No. 1. pp.9–24.